\documentclass[twocolumn]{revtex4}
\usepackage[dvips]{graphicx}
\usepackage{bm}
\usepackage{amsmath}

\begin{document}

\title{Effect of weak disorder on delocalization properties of gapped graphene superlattices}

\author{E.S.~Azarova}
\author{G.M.~Maksimova}
\altaffiliation[Corresponding author. ]{Tel.: +7 831 4623304; \\
E-mail address: maksimova.galina@mail.ru (G.M. Maksimova)}

\address{Department of Theoretical Physics, University of Nizhny Novgorod, 23 Gagarin Avenue, 603950 Nizhny Novgorod, Russian Federation}

\begin{abstract}

We study the effect of weak disorder on the delocalization properties of gapped graphene superlattice (SL) formed by periodically located rectangular potential barriers. We consider two types of the SLs: the SLs with uniform gap and SLs consisting of alternating layers of gapped and gapless graphene regions. Using the perturbative approach we obtain an analytical expression for the inverse localization length (ILL) derived for the case of randomly fluctuating \textit{geometric} and \textit{energetic} parameters. In the first case, when the barrier (well) width fluctuates around its mean value, the corresponding equation for the ILL reveals the presence of the Fabry-Perot resonances, at which the localization length diverges. These resonances are exact, i.e., are stored in any degree of disorder. It has been found that the localization properties manifest stronger for the particles with energies lying in the non-resonant bands where our approach is extremely sensitive to the degree of disorder. For the case of weakly fluctuating both barrier and well widths we analytically obtain ILL taking correlations into account. The main effect of the correlations, which lead to an increase (or decrease) in the localization length, was revealed near the double resonance arising at coincidence of two Fabry-Perot resonances associated with barrier and well widths. The random fluctuations of the potential strength also lead to the delocalization resonances. However, they exist only in a weak-disorder approximation. We found that, for an array composed of alternating strips of gapless and gapped graphene modifications these resonances can appear only for normally incident particles in contrast to the SL with a uniform gap. For such particles, the delocalization resonances occur also in the purely random potential. This means, in particular, that in the one-dimensional case, not all the states of the massive Dirac particles are localized in the presence of weak disorder.

\end{abstract}

\maketitle

\section{Introduction}

Recent years, much attention of both theoreticians and experimentalists has been paid to the graphene-based superlattices  (SLs)~\cite{Park, Brey, Barb, Dell, Le, Arov, Esm1, Dub, Wang}
. Such interest results from the prediction of possible engineering the system band structure by the periodic potential. This opens different ways to fabricate graphene-based electronic devices. An undoped graphene is a zero-gap semiconductor. This property leads, in particular, to the total transparency of any potential barrier for normally incident electrons (an analog of the Klein paradox). At the same time, most electronic applications are based on the presence of a gap between valence and conduction bands. Therefore, it is crucial to induce a band gap in Dirac points to control the transport of carriers. For this purpose, several approaches have been studied both theoretically and experimentally. Among them, size quantization in armchair nanoribbons, as well as application of external electric potentials along the sample edges in zigzag nanoribbons were considered~\cite{Son, Han, Yan, Apel}
. It has been shown that the gap value increases with decreasing the nanoribbon width and strongly depends on the detailed structure of the ribbon edges. Other proposed mechanisms, which are effective also in broad graphene sheets, are strain-induced gap opening~\cite{Gui, Pereira}
, chemical effects of adsorbent atoms and molecules~\cite{Rib} 
and substrate-induced band gap formation owing to a breaking the sub-lattice symmetry~\cite{Giov}
. The energy spectrum of the Dirac electrons in an epitaxially grown on a SiC substrate graphene layer has been measured by Zhou et al.~\cite{Zhou}. 
 They observed an opened up energy gap of about 260 meV in the electronic spectrum. It is worth noting also that the $\vec{k}\cdot\vec{p}$ Hamiltonian of other two-dimensional materials with hexagonal symmetry, such as molybdenum disulfide (MoS$_2$), is similar to the Dirac Hamiltonian for massive particles~\cite{Xiao, Li}
.

Besides, in some publications, various SLs based on graphene with spatially inhomogeneous gap (i.e., the particle’s mass), and the possibility of their creation are discussed~\cite{Peres, Gomes, Giav, Ratn, Maks, Azar}
. It was shown, that the spatial mass dependence leads to the suppression of Klein tunneling and induces confined states~\cite{Peres, Giav}
. One way of making graphene heterostructures with the required gap modulation is a deposition of graphene on an inhomogeneous substrate fabricated from different dielectrics. It is also possible to use for these purpose an inhomogeneously hydrogenated graphene or graphene sheet with nonuniformly deposited CrO$_3$ molecules.

In our previous works~\cite{Maks, Azar} 
we investigated the electronic band structure and transport properties of graphene superlattice in which the gap and potential profile are piecewise constant functions. It was shown that in such SL, up to some critical value of potential $V_c$ , allowed subbands are separated by gaps. At $V>V_c$ the contact or cone-like Dirac points appear in the spectrum. It was also found that each a new Dirac point manifests itself as a conductivity resonance and a narrow dip in the Fano factor F similarly to a gapless SL. However between the resonances, behavior of the Fano factor in the considered structure is more complicated and differs from pseudo-diffusive behavior ($F=1/3$) typical for a gapless SL~\cite{Fert}
.

Meanwhile, real graphene superlattices cannot be perfectly periodic due to random imperfections resulting, for example, from variations of the system parameters such as potential height, gap value, potential width or barrier spacing. It is well known, that in the presence of white-noise disorder all the electronic states are localized in the thermodynamic limit for a traditional semiconductor superlattice. On the contrary, a sample of gapless graphene in the presence of a random one-dimensional potential becomes completely transparent for the normally incident particles regardless of the sample length and strength of disorder. This means, that the states of the massless Dirac particles are entirely delocalized for arbitrary disorder strength due to the chiral symmetry~\cite{Nom, Zhu}
. The transport properties of disordered graphene superlattices have been studied by several groups~\cite{Bliokh, Abed, Zhao}
. It was found that the transport and spectral properties of gapless graphene superlattices created by applying either periodic or disordered smooth scalar potentials are strongly anisotropic. The dc conductance of graphene superlattice consisting of p-n junctions for various strengths of structural disorder imposed on the material has been investigated numerically in Ref.~\cite{Abed}
. It was shown that there exists a range of angles around the normal incidence angle, for which the transmission becomes finite in the presence of structural white-noise disorder. For weakly disordered both scalar-potential and vector-potential graphene SLs the localization behavior of massless Dirac particles was studied in Ref.~\cite{Zhao} 
numerically as well as analytically by a weak-disorder expansion. In particular, strong dependence of the Lyapunov exponent (the inverse localization length) on the incident angle of the charge carriers injected to a graphene superlattice has been predicted. The effects of gap fluctuations on transmission and conductance of the monolayer and bilayer graphene SLs were treated numerically in Ref.~\cite{Esm2}
.

The aim of this work is to study the effect of weak disorder on the localization length and transport properties of disordered gapped graphene SLs including the samples with spatially inhomogeneous gap. We extend the theoretical study developed earlier for the case of gapless graphene SLs~\cite{Zhao} 
and obtain an analytical expression for the localization length derived for the cases of randomly fluctuating parameters of the SLs. This expression is in a good agreement with direct simulations. We also take into account possible correlations for the case of weakly fluctuating widths of layers forming the unit cell of the superlattice.

The paper is organized as follows. Section~II is devoted to the description of the model and the method. In Sec.~III we present the dispersion relation and transmission for graphene-based multibarrier periodic structure with spatially dependent gap in the presence of the step-like potential. The analytical expression for the inverse localization length of periodic-on-average disordered graphene SLs as well as the results of numerical simulations are presented in Sec.~IV. We make a summary and concluding remarks in Sec.~V.

\section{Model and method}

\begin{figure}[t] \centering
\includegraphics*[scale=0.6]{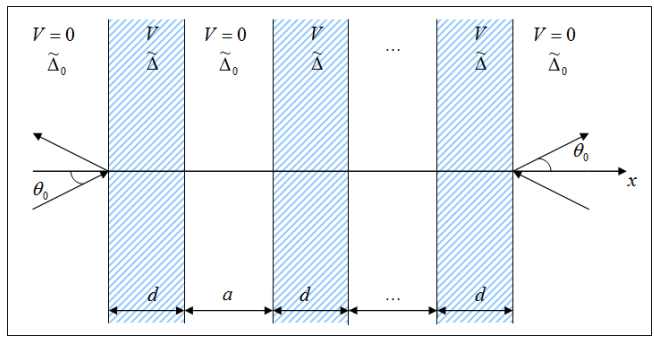}
\caption{Models of a periodic structure: $\tilde{\Delta}=\tilde{\Delta}_0$ -- for the homogeneous superlattice (HSL); $\tilde{\Delta}_0=0$ -- for the superlattice formed by the alternating layers of gapless graphene (MSL).} \label{fig1}
\end{figure}

We consider the propagation of an electron through the lateral disordered structure formed by a sequence of $N$ barrier regions with width $d_n(n=1,2,\dots N)$ separated by inter-barrir distance $a_n$ (wells), as shown in Fig.~\ref{fig1}. The disorder is introduced as random, small variations of the barrier strengths or other barrier characteristics (e.g., gap value in gapped graphene) as well as the barrier and well thicknesses around their mean values. In the absence of the disorder the considered system is periodic with the period $l=a+d$, with $d=\langle d_n\rangle$, $a=\langle a_n\rangle$. The main subject of our study is the localization length $L_{loc}$, defined as
\begin{eqnarray}\label{eq1}
\gamma=\frac{l}{L_{loc}}=-\lim_{N\to\infty}\left\langle \frac{\ln T_N}{2N}\right\rangle,
\end{eqnarray}
where $T_N$ is the random tramsmission coefficient of a sample of the length $Nl$ and angular brackets are used to denote averaging over different disorder realizations. To calculate the transmission coefficient, we use the common transfer-matrix approach.

In the general case, between barriars, where $V(x)=0$, the wave function $\psi_n(x,y)$ can be written as $\psi_n(x,y)=e^{ik_yy}\psi_n(x)$, where $\psi_n(x)$ is a superposition of the wave functions of right- and left-moving particles
\begin{eqnarray}\label{eq2}
\psi_n(x)=A_ne^{ik_xx}u+B_ne^{-ik_xx}\upsilon, \text{ } x_n+d_n\le x\le x_{n+1}
\end{eqnarray}
and $u$, $\upsilon$ are spinor amplitudes, defined by the specific Hamiltonian. The amplitudes $\psi_n^+=A_ne^{ik_xx_n}$ and $\psi_n^-=B_ne^{-ik_xx_n}$ in neighboring wells are mapped from $n$ to $n+1$ by the transfer matrix for a single unit
\begin{eqnarray}\label{eq3}
\begin{pmatrix}
\psi_{n+1}^+ \\   
\psi_{n+1}^-   
\end{pmatrix}
=\hat{M}_n
\begin{pmatrix}
\psi_{n}^+ \\   
\psi_{n}^-   
\end{pmatrix}
\end{eqnarray}
with
\begin{eqnarray}\label{eq4}
\hat{M}_n=
\begin{pmatrix}
\frac{e^{ik_xa_n}}{t_n^*} & \frac{-r_n^*e^{ik_xa_n}}{t_n^*} \\   
\frac{-r_ne^{-ik_xa_n}}{t_n} & \frac{e^{-ik_xa_n}}{t_n}    
\end{pmatrix}
.
\end{eqnarray}
Here reflection $(r_n)$ and transmission $(t_n)$ amplitudes are determined by the parameters of the $n$th barrier, as well as dynamic characteristics of the particles, i.e., energy $E$ and momentum $\hbar k_y$. By construction, the transfer matrix across $N$ barriers is the product
\begin{eqnarray}\label{eq5}
\hat{P}_N=\prod_{n=1}^N{\hat{M}_n},
\end{eqnarray}
so the transmission probability
\begin{eqnarray}\label{eq6}
T_N=\left|(\hat{P}_N)_{11}\right|^{-2}.
\end{eqnarray}

For infinite periodic structure with period $l=a+d$ matrix elements $\hat{M}_n$ does not depend on $n$ and the electronic band structure of the corresponding superlattice is governed by the following relation
\begin{eqnarray}\label{eq7}
2\cos{Kl}=Tr\hat{M}
\end{eqnarray}
with $K$ the Bloch wave vector. As known (see, e.g., Ref.~\cite{Mark}
), the transmission coefficient $T_N^0$ of the array of $N$ identical cells (Fig.~\ref{fig1}
) can be written in closed form as follows
\begin{eqnarray}\label{eq8}
T_N^0=\left(1+\left|\frac{r}{t}\right|^2\left(\frac{\sin{N\eta}}{\sin\eta}\right)^2\right)^{-2},
\end{eqnarray}
where the Bloch phase $\eta=Kl$ is defined by Eq.~(\ref{eq7}). Note, that a similar expression holds for the transmission coefficient ${T'}_N^0$ through an array of length $L=Nl$ bounded by the regions, characterized by barrier parameters. In this case ${T'}_N^0$ depends on the reflection ($r'$) and transmission ($t'$) amplitudes across the \textit{inter-barrier region (well)}.

In the presence of weak disorder the localization length can be found  by substituting Eqs.(\ref{eq4}) -- (\ref{eq6}) into definition (\ref{eq1}) and expanding the logarithm up to quadratic terms in disorder. To perform these calculations, we need to know the transmission and reflection amplitudes for a single barrier.

\section{Gapped graphene superlattices}

System under consideration is graphene-based multibarrier structure with spatially dependent gap $\tilde\Delta(x)$ in the presence of the step-like potential $V(x)$. For a periodic system $V(x)=V(x+l)$, $\tilde\Delta(x)=\tilde\Delta(x+l)$, where
\begin{eqnarray}\label{eq9}
V(x),\tilde\Delta(x)=
\begin{cases}
V,\tilde\Delta & \text{for $0<x<d$,}\\
0,\tilde\Delta_0&\text{for $d<x<l$.}
\end{cases}
\end{eqnarray}
The wave functions of charged particles in such model obey the Dirac equation with the Hamiltonian
\begin{eqnarray}\label{eq10}
\hat{H}=\upsilon_F\hat{\mathbf{p}}\sigma+\tilde{\Delta}(x)\sigma_z+V(x),
\end{eqnarray}
with $\hat{\mathbf{p}}$ -- the momentum operator, $\sigma=(\sigma_x,\sigma_y)$, $\sigma_z$ the Pauli matrices and $\upsilon_F\approx 10^6$~ms$^{-1}$ the Fermi velocity. The equation $(\hat{H}-E)\psi=0$ admits the plain wave solution of the form $\psi (x,y)=e^{ik_yy}\psi (x)$ with
\begin{eqnarray}\label{eq11}
&&\psi (x)=Ae^{iQx}
\begin{pmatrix}
1 \\
\frac{Q+ik_y}{\varepsilon-\upsilon(x)+\Delta(x)}
\end{pmatrix}\nonumber\\
&&+Be^{-iQx}
\begin{pmatrix}
1 \\
\frac{-Q+ik_y}{\varepsilon-\upsilon(x)+\Delta(x)}
\end{pmatrix}
.
\end{eqnarray}
For convenience, hereafter all lengths will be expressed in the units of the mean period $l$. The natural energy scale is $E_{SL}=\frac{\hbar\upsilon_F}{l}$ ($E_{SL}\approx6.25$ meV for $l=100$ nm), so that other dimensionless parameters are $\varepsilon=E/E_{SL}$, $\upsilon(x)=V(x)/E_{SL}$, $\Delta(x)=\tilde{\Delta}(x)/E_{SL}$,  where $V(x)$ and $\tilde{\Delta}(x)$ in the barrier and well regions are determined by Eq.~(\ref{eq9}). $Q$ is the dimensionless wave vector along the $x$-axis
\begin{eqnarray}\label{eq12}
Q=
\begin{cases}
q_x & \text{for barrier,}\\
k_x &\text{for well},
\end{cases}
\end{eqnarray}
where
\begin{eqnarray}\label{eq13}
q_x=\sqrt{(\varepsilon-\upsilon)^2-\Delta^2-k_y^2}, \text{ } k_x=\sqrt{\varepsilon^2-\Delta_0^2-k_y^2}.
\end{eqnarray}
By applying the continuity of the wave function at the boundaries, we obtain the transmission ($t$) and reflection ($r$) amplitudes for electrons incident at an angle $\theta_0$ with respect to the $x$-axis (Fig.~\ref{fig1})
\begin{eqnarray}\label{eq14}
\frac{1}{t}=\cos\beta+i\frac{\varepsilon\upsilon+\Delta\Delta_0-\varepsilon^2+k_y^2}{k_xq_x}\sin\beta,
\end{eqnarray}
\begin{eqnarray}\label{eq15}
&& \frac{r}{t}=-i\sin\beta e^{i\theta_0}\frac{\left(\varepsilon\left(\Delta-\Delta_0\right)+\upsilon\Delta_0\right)k_x}{kk_xq_x}\nonumber\\ &&\frac{+i\left(\Delta_0^2-\Delta\Delta_0-\varepsilon\upsilon\right)k_y}{kk_xq_x},
\end{eqnarray}
where $\beta=q_xd$, $k=\sqrt{\varepsilon^2-\Delta_0^2}$ the particle wave vector outside the barrier, $\theta_0=\tan^{-1}\frac{k_y}{k_x}$. Then tunneling through a single \textit{barrier} is given by
\begin{eqnarray}\label{eq16}
T\left(\varepsilon,k_y\right)=\left|t\right|^2=\left(1+\left(f^2\left(\varepsilon,k_y\right)-1\right)\sin^2\beta\right)^{-1}
\end{eqnarray}
with
\begin{eqnarray}\label{eq17}
f\left(\varepsilon,k_y\right)=\frac{\varepsilon\upsilon+\Delta\Delta_0-\varepsilon^2+k_y^2}{k_xq_x}.
\end{eqnarray}
Transmittance through a single \textit{well} differs from this expression only by replacing $\beta\to\alpha=k_xa$. At $\Delta=\Delta_0=0$ the expression (\ref{eq16}) coincides with the similar to gapless graphene~\cite{Kats}
, and for $\Delta_0=0$, $\Delta\ne0$ is the same as that established in Ref.~\cite{Azar}
. Using Eq.~(\ref{eq8}), we obtain the transmission across $N$ identical \textit{barriers}
\begin{eqnarray}\label{eq18}
&&T_N^0\left(\varepsilon,k_y\right)=\Biggl(1+\left(f^2\left(\varepsilon,k_y\right)-1\right)\sin^2\beta\nonumber\\
&&\cdot\left(\frac{\sin N\eta}{\sin\eta}\right)^2\Biggl)^{-1},
\end{eqnarray}
where Bloch phase $\eta$ according to Eqs.~(\ref{eq4}), (\ref{eq7}), (\ref{eq14}) can be obtained from dispersion relation
\begin{eqnarray}\label{eq19}
\cos\eta=\cos\alpha\cos\beta+f\left(\varepsilon,k_y\right)\sin\alpha\sin\beta.
\end{eqnarray}
As follows from the expression~(\ref{eq18}) the transmission $T_N^0\left(\varepsilon,k_y\right)=1$ for any $N$ under conditions $\sin\beta=0$, $\beta\ne0$ or when $\sin N\eta/\sin\eta=0$. The first equation $q_xd=\pi m$ determines the Fabry-Perot resonances~\cite{Milt, Shytov, Masir} related with the barrier regions and the second produces $N-1$ Fabry-Perot oscillations in each allowed energy band. Similarly, for particles incident on array of $N$ unit cells from the barrier region ${T'}_N^0\left(\varepsilon,k_y\right)=1$ when $\sin\alpha=0$, $\alpha\ne0$ or $\sin N\eta/\sin\eta=0$. Note also, that for gapless superlattices with $\Delta=\Delta_0=0$, gapped with a uniform gap (homogeneous or HSL) with $\Delta=\Delta_0\ne0$ and for the superlattices, formed by alternating strips of gapless and gapped graphenes (mixed or MSL) with $\Delta_0=0$, $\Delta\ne0$ the function $f\left(\varepsilon,k_y\right)$ (\ref{eq17})determining transport properties and the energy spectrum of the SLs has the same form
\begin{eqnarray}\label{eq20}
f\left(\varepsilon,k_y\right)=\frac{\varepsilon\upsilon-k_x^2}{k_xq_x},
\end{eqnarray}
where wave vectors $k_x$ and $q_x$ are defined by formula~(\ref{eq13}) for each type of the superlattice.

\section{Localization length for disordered graphene structures}

In what follows, we consider the disordered multibarrier graphene structures in which the disorder is caused by random fluctuations of barrier strength, or gap magnitude (inside the barriers) as well as by random variations of both barrier and well widths. Specifically, we assume a weakness of both types of disorder
\begin{eqnarray}\label{eq21}
s_n=s\left(1+\rho_n^s\right), \text{ } s=\upsilon,\Delta \text{ or } d,a.
\end{eqnarray}
Here the index $n$ enumerates the $n$th unit ($d,a$) cell, $\rho_n^s$ is random \textit{uncorrelated} variables with zero average and small variances $\sigma_s^2\ll1$, i.e.
\begin{eqnarray}\label{eq22}
\left\langle\rho_n^s\right\rangle=0, \text{ } \left\langle\rho_n^s\rho_{n'}^{s'}\right\rangle=\sigma_s^2\delta_{nn'}\delta_{ss'}.
\end{eqnarray}
The averaging $\langle\dots\rangle$ is performed over the whole array of layers or due to the ensemble averaging, that is equivalent to the assumption. Numerically, for generating random sequences $\rho_n^s$ we use the flat distribution on a finite interval $[-\delta,\delta]$. An analytical expression for the inverse localization length (ILL) $\gamma$ can be obtained by the method of perturbation theory. To do this, follow Zhao et.al.~\cite{Zhao}
, represent the expression for the transfer matrix for a single unit~(\ref{eq4}) as
\begin{eqnarray}\label{eq23}
M_n=
\begin{pmatrix}
e^{im_n}\sec\varphi_n & e^{ip_n}\tan\varphi_n \\
e^{-ip_n}\tan\varphi_n & e^{-im_n}\sec\varphi_n
\end{pmatrix}
,
\end{eqnarray}
where $\sin\varphi_n=|r_n|$, $\cos\varphi_n=|t_n|$ and the parameters $m_n$ and $p_n$ are determined by the value of $k_xa$ and the phases of the reflected and transmitted waves (Eq.~(\ref{eq4})). Then the weak-disorder ILL or Lyapunov exponent depends only on the parameters of the underlying \textit{regular} array
\begin{eqnarray}\label{eq24}
\gamma_s=\frac{s^2\sigma_s^2}{2}\tan^2\varphi\left[{p'}^2+{\left(\frac{\sin m}{\sin\varphi}\right)'}^2\frac{\tan^2\varphi}{\sin^2\eta}\right],
\end{eqnarray}
where the prime $(\dots)'$ denotes differentiation with respect to the perturbation variable $s$. Using Eqs.~(\ref{eq14}), (\ref{eq15}) it is not difficult to show that for the considered gapped SLs
\begin{eqnarray}\label{eq25}
\tan^2\varphi=\left(f^2\left(\varepsilon,k_y\right)-1\right)\sin^2\beta,
\end{eqnarray}
\begin{eqnarray}\label{eq26}
\frac{\sin m}{\sin\varphi}=\frac{\sin\alpha\cot\beta-f\left(\varepsilon,k_y\right)\cos\alpha}{\sqrt{f^2\left(\varepsilon,k_y\right)-1}},
\end{eqnarray}
\begin{eqnarray}\label{eq27}
p=\alpha-\frac{\pi}{2}-\theta_0-\xi,
\end{eqnarray}
\begin{eqnarray}\label{eq28}
\xi=\tan^{-1}\left(\frac{\Delta_0^2-\Delta\Delta_0-\varepsilon\upsilon}{\varepsilon(\Delta-\Delta_0)+\upsilon\Delta_0}\tan\theta_0\right).
\end{eqnarray}
Note that Eq.~(\ref{eq24}) is correct inside the energy bands $(TrM<2)$ apart from the band edges $\eta=0,\pi$. When $TrM>2$, the energy lies in the forbidden miniband. In this case ILL $\gamma_s$ is defined by $\lambda_+$, the largest of two eigenvalues of the transfer matrix $\hat{M}$ $\gamma_s=\ln|\lambda_+|$.

Using the weak-disorder approach it is possible to generalize the expression for ILL~(\ref{eq24}) to the case when the correlations of fluctuating quantities (e.g., the geometric parameters of the structure) exist. Details of the calculation are given in the Appendix.

\subsection{Geometric disorder}

\begin{figure*}[t] \centering
\includegraphics*[scale=1.2]{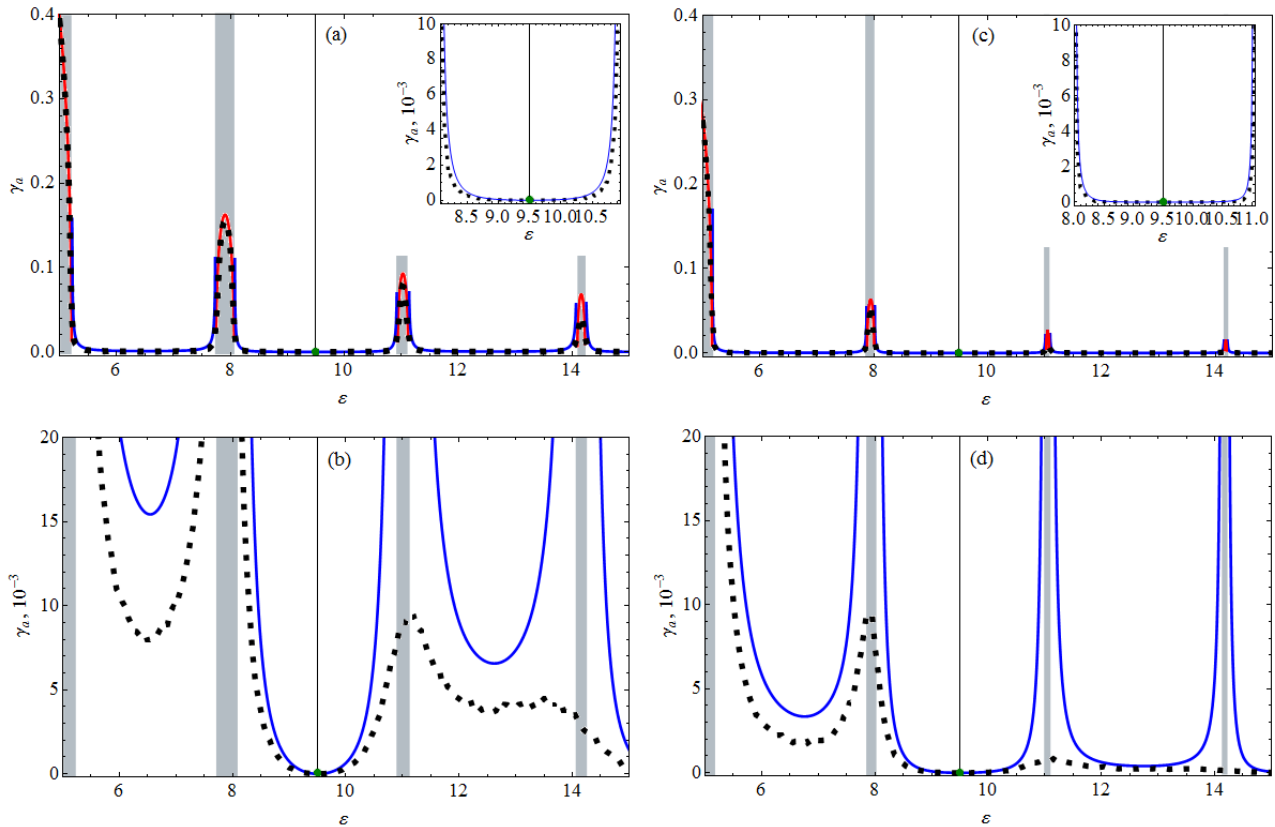}
\caption{Inverse localization length $\gamma_a$ versus particle energy $\varepsilon$ at $k_y=0$ for graphene MSL (a, b) and HSL (c, d) corresponding to the fluctuation of distance between the barrier with two disorder strength: $\delta=0.05$ (a), (c) and $\delta=0.2$ (b), (d). For all cases $a=d=0.5$, $\upsilon=\pi$, $\Delta=\pi/3$. The continuous (blue) curve corresponds to the analytical results [Eqs.~(\ref{eq32}), (\ref{eq33})], dotted line presents numerical data for the structure length $N=10^3$ with an ensemble averaging performed over $80$ realizations of disorder. The continuous (red) line shows the dependence of ILL on $\varepsilon$ in the forbidden minibands, shaded in Figure. Vertical lines mark the delocalization resonance positions, placed in the resonance zone shown in the insets.} \label{fig2}
\end{figure*}
For weakly fluctuating widths of layers (positional or geometric disorder), according to Eqs.~(\ref{eq24}), (\ref{eqa9}) the Lyapunov exponent, which includes the correlation term, can be written as
\begin{eqnarray}\label{eq29}
&& \gamma_{a,d}=\left(f^2\left(\varepsilon,\theta_0\right)-1\right)\frac{\bigl[\alpha^2\sigma_a^2\sin^2\beta+\beta^2\sigma_d^2\sin^2\alpha}{2\sin^2\eta}\nonumber\\
&& \frac{-2\alpha\beta\sigma_{ad}\sin\alpha\sin\beta\cos\eta\bigl]}{2\sin^2\eta},
\end{eqnarray}
where $f\left(\varepsilon,k_y\right)$ is given by Eq.~(\ref{eq20}) and two types of the considered gapped graphene structures differ only in the value of the wave vector $k_x$ in well region $k_x=k_x^{HSL}=\sqrt{\varepsilon^2-\Delta^2-k_y^2}$ for the SL with uniform gap and $k_x=k_x^{MSL}=\sqrt{\varepsilon^2-k_y^2}$ for the SL with piecewise constant gap.

Assume that only the distances between the barriers display random fluctuations around their mean value $a$, that is $\sigma_d^2=\sigma_{ad}=0$, $\sigma_a^2=\delta^2/3$ and $\delta$ determines the degree of disorder. In this case Eq.~(\ref{eq29}) indicates that the localization length turns into infinity when performing the Fabry-Perot resonance conditions $q_xd=\pi m$. Moreover, this result is exact, that is valid for any degree of disorder. Indeed, in this case transmission amplitude across a single barrier $t=1$ (\ref{eq14}) resulting in total transparency of the array from $N$ identical barriers located randomly.
Figure~\ref{fig2} compares the analytical results for $\gamma_a^{MSL}$ and $\gamma_a^{HSL}$ (left and right column, respectively) with the numerical data, for a varying particle energy $\varepsilon$, fixed $\theta_0=0$ and for two different values of disorder: $\delta=0.05$ (Figs.~\ref{fig2}(a), \ref{fig2}(c)) and $\delta=0.2$ (Figs.~\ref{fig2}(b), \ref{fig2}(d)). The potential strength is $\upsilon=\pi$ and the gap value is $\Delta=\pi/3$. Numerical data are obtained for an array size $N=10^3$ with additional average over $80$ realization. From Figs.~\ref{fig2}(a), \ref{fig2}(c) it can be seen that the theoretical expressions for the inverse localization length for both models of the SL provide a very good description for the case of weak disorder ($\delta=0.05$) for this range of energies. For higher disorder strength ($\delta=0.2$) the agreement between the analytical and numerical calculations remains good for the SL with a uniform gap at least for energies $\varepsilon>8$ (Fig.~\ref{fig2}(d)). But the lattice formed by alternating stripes of the gapped and gapless graphene (MSL), is more sensitive to fluctuations of the inter-barrier distance (Fig.~\ref{fig2}(b)). It is also clearly seen that in the neighborhood of the Fabry-Perot resonances the localization is strongly suppressed and the greater the amount of the disorder, the narrower this neighborhood. These results are in  complete correspondence with those  obtained (and has been confirmed experimentally) in Ref.~\cite{Luna}
, which deals with the propagation of electromagnetic waves through one-dimensional disordered bi-layer  structures whose unit cell consists of two different dielectrics. Applicability approximate expressions (\ref{eq29}) at a given disorder strength depends on of the lattice parameters. Thus, with increasing the gap magnitude   visible discrepancies with the results of numerical calculations are observed even at the disorder strength $\delta=0.05$.

\begin{figure}[t] \centering
\includegraphics*[scale=0.6]{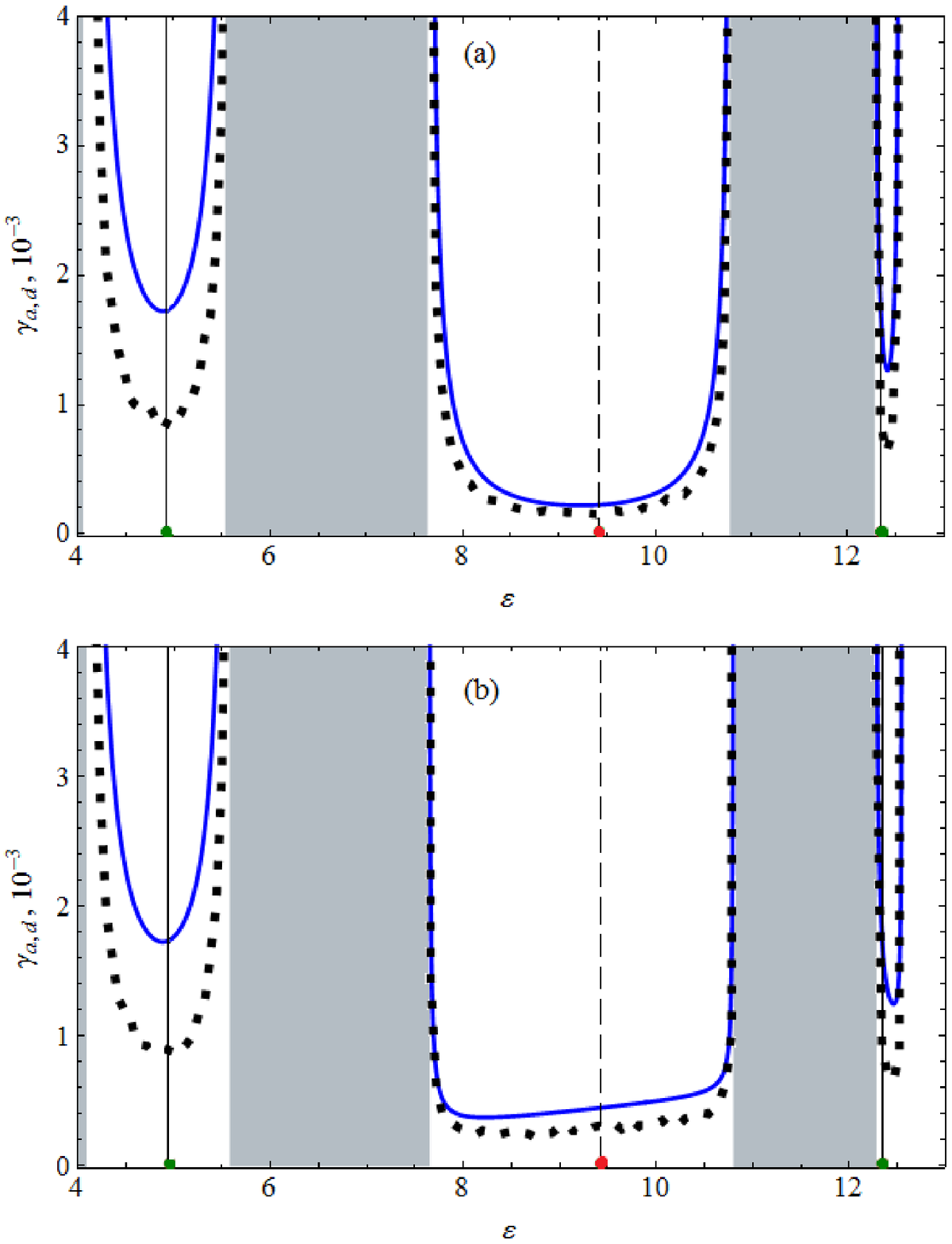}
\label{fig3}
\end{figure}
\begin{figure}[t] \centering
\includegraphics*[scale=0.6]{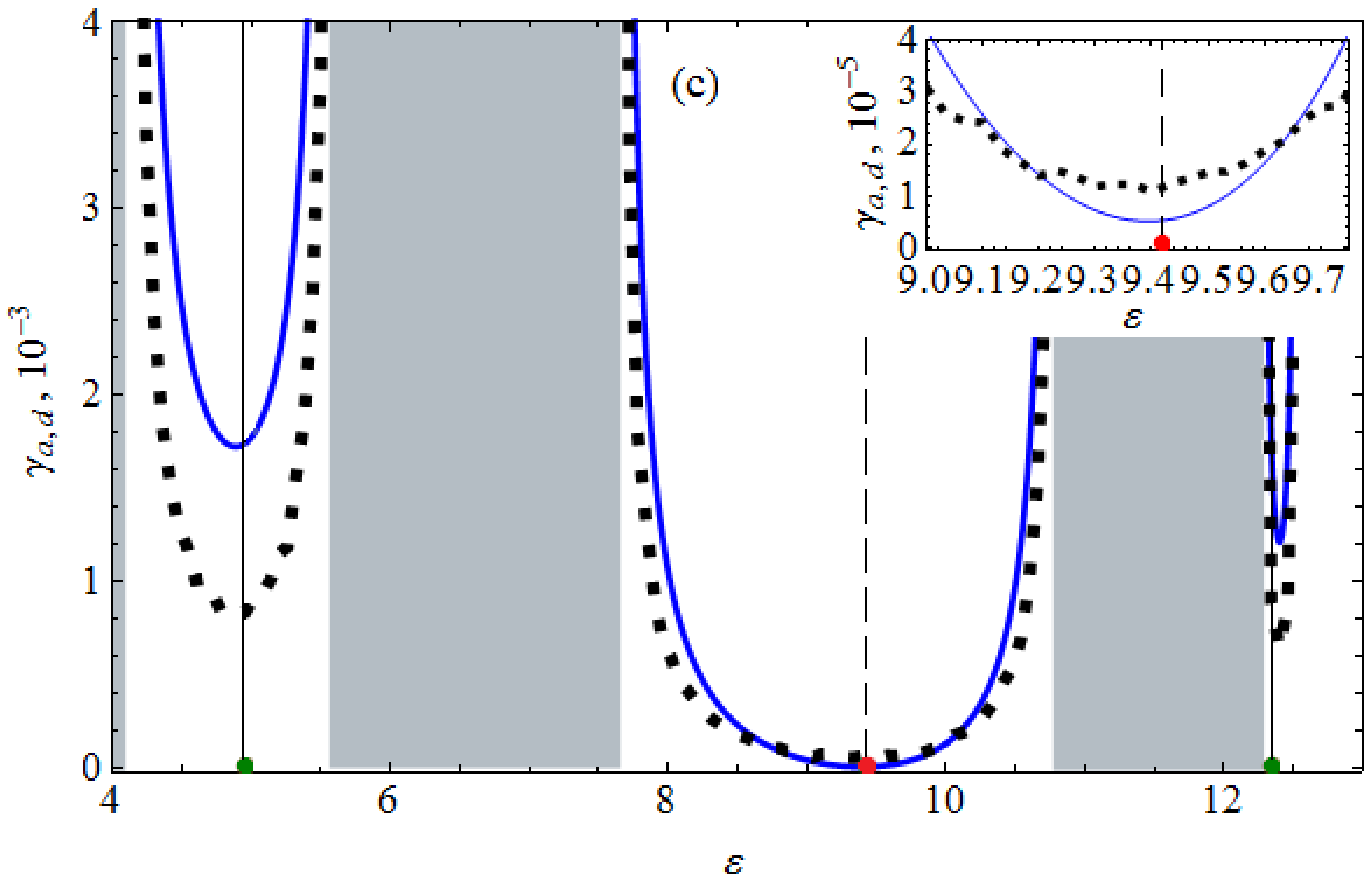}
\caption{Inverse localization length $\gamma_{a,d}$ versus particle energy $\varepsilon$ for uncorrelated (a), completely correlated (b), and anticorrelated (c) disorder at $\theta_0=\cos^{-1}2/3$ for graphene MSL with $\upsilon=8\pi$, $\Delta=2\pi$ corresponding to the fluctuations both the distance between the barriers and their width near an average value $a=d=0.5$ with disorder strength $\delta=0.005$. The solid line represents the analytical results, dotted line corresponds to the numerical simulations for an array of $N=5\cdot10^3$ ($N=3\cdot10^4$ for insert) with the average over $100$ realizations of disorder. The vertical solid lines indicate the positions of the Fabry-Perot resonances arising when $q_xd=\pi;3\pi$. The vertical dashed lines depict the positions of the double resonances that occur when $q_xd=2\pi$ and $k_xa=\pi$. Insert illustrates the behavior of $\gamma_{a,d}^{MSL}$ in a neighborhood of double resonance.} \label{fig3}
\end{figure}
Similarly, for fluctuating barrier widths ($\sigma_a^2=\sigma_{ad}=0$, $\sigma_d^2=\delta^2/3$) the expression~(\ref{eq29}) for the inverse localization length $\gamma_d$ reveals the delocalization resonances for both types of the SLs under the conditions $\sin\alpha=0$. When these conditions are met, the transmittance ${T'}_N$ through $N$ identical \textit{wells}, separated by barriers, the widths of which vary randomly, is equal to one for any $N$. On the other hand, the transmission coefficient $T_N$ through $N$ regularly spaced \textit{barriers} with fluctuating width does not decrease exponentially with increasing $N$, which leads to $\gamma_d=0$, i.e. the suppression of localization.

For the array with randomly varying both barrier width and inter-barrier spacing the ILL $\gamma_{a,d}$ is obtained from the general expression~(\ref{eq29}). As above, we take $a=d$. Then for uniform random perturbation with the same amplitudes on both layers, we have $\sigma_a^2=\sigma_d^2=\delta^2/3$. If the disorder is uncorrelated, then $\sigma_{ad}=0$. For completely correlated disorder, when the barrier and well widths in the period change in the same way, we have $\sigma_{ad}=\delta^2/3$. In the case of completely anticorrelated disorder period of the SL remains constant and we take $\sigma_{ad}=-\delta^2/3$~\cite{Mog}
. For the Fabry-Perot resonances occurring when $\alpha=\pi n$ or $\beta=\pi k$ with $n,k=1;2;3;\dots$, the factor $\sin\alpha$ or $\sin\beta$ in Eq.~(\ref{eq29}) vanishes, so that the correlations do not affect the localization properties of the structure. It may happen that some of the resonances due to different graphene layers coincide for certain values of $\varepsilon$ and $\theta_0$ that it is possible under the condition $\sin\alpha=0$, $\sin\beta=0$. But under these conditions the denominator $\gamma_{a,d}$ vanishes also (\ref{eq19}), so localization length remains finite and its value significantly depends on the existing correlations. In Fig.~\ref{fig3} the ILL for: (a) uncorrelated, (b) completely correlated and (c) anticorrelated disorder and oblique incidence is shown as a function of energy for disorder strength $\delta=0.005$ for graphene MSL. In this Figure the central band includes the value of the energy $\varepsilon_0=3\pi$ which determine the position of the double resonance defined by the conditions $\alpha=\pi$, $\beta=2\pi$. Fabry-Perot resonances corresponding to the conditions $\beta=\pi$ and $\beta=3\pi$, for the chosen lattice parameters occur in the first and third allowed energy bands. Analysis~(\ref{eq29}) shows that the values of $\gamma_{a,d}^{MSL}$ corresponding to such resonances close to the minimum values of the ILL in the relevant energy bands, but do not coincide with them. It is interesting also to note that for the SLs with spatially inhomogeneous gap the resonance values of the ILL does not depend on the energy
\begin{eqnarray}\label{eq30}
\gamma_{a,d}^{MSL}\left(\beta=\pi k\right)=\frac{\left(\Delta^2+\upsilon^2\tan^2\theta_0\right)d^2\sigma_d^2}{2}.
\end{eqnarray}
\begin{figure*}[t] \centering
\includegraphics*[scale=1.2]{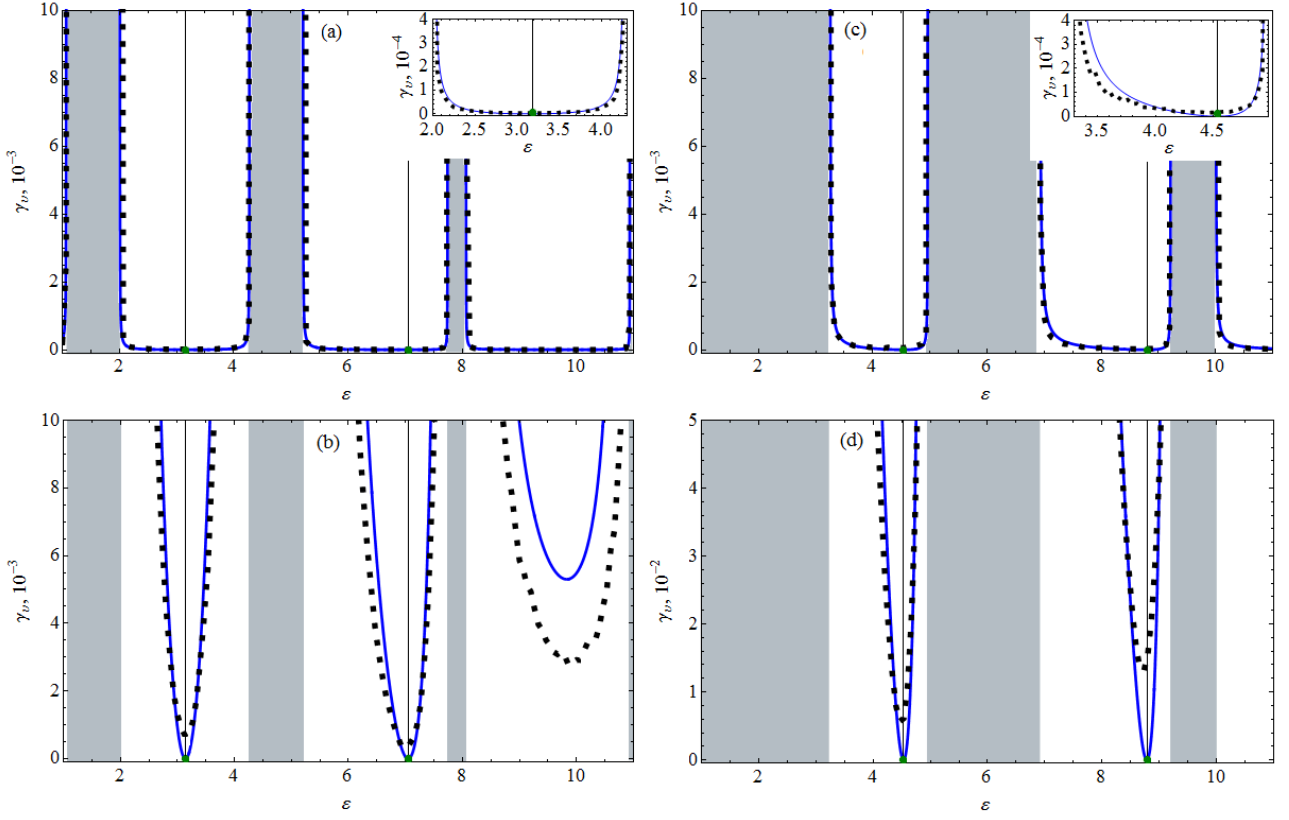}
\caption{Inverse localization length $\gamma_{\upsilon}$ as a function of energy $\varepsilon$ for graphene MSL for an angle of incidence $\theta_0=0$ (a, b) and HSL for $\theta_0=\pi/6$ (c, d) with fluctuating barrier height for two disorder strength: $\delta=0.01$ (a, c); $\delta=0.5$ (b, d). For both cases $a=d=1/2$, $\upsilon=\pi$, $\Delta=\pi/3$, the other notations are the same as in Figure~\ref{fig2}. Numerical calculations were performed for an array composed of $N=5\cdot10^3$ until cells ($N=2\cdot10^4$ for inserts).} \label{fig4}
\end{figure*}
Fig.~\ref{fig3} clearly demonstrates a significant discrepancy of approximate analytical calculations with numerical data in the first and third energy bands, even with such a relatively small amount of disorder, as $\delta=0.005$. In the central energy band, containing double resonance, our analytical predictions, based on the formula~(\ref{eq29}), are more consistent with the numerical data. Clearly seen, that the most influence of correlations appear close to the double resonance: completely correlated disorder suppresses the localization length (Fig.~\ref{fig3}(b)) and anticorrelated disorder term in Eq.~(\ref{eq29}) leads to its increasing (Fig.~\ref{fig3}(c)). For the superlatice with uniform gap the results are similar.

\subsection{Compositional disorder}

In the case when the barrier height fluctuates around their mean value the Lyapunov exponent $\gamma$ given by Eq.~(\ref{eq24}) for both types of superlattices is defined by the expressions
\begin{eqnarray}\label{eq31}
&& \gamma_\upsilon^{MSL}=\frac{\upsilon^2\sigma_\upsilon^2}{2q_x^2}\nonumber\\
&& \cdot\left[\frac{k_y^2\Delta^2\sin^2\beta}{S}+\frac{S\left(F^M\sin\alpha+G^M\cos\alpha\right)^2}{k_x^2q_x^2\sin^2\eta}\right],
\end{eqnarray}
where
\begin{eqnarray}\label{eq32}
S=\upsilon^2k_y^2+\Delta^2k_x^2,
\end{eqnarray}
\begin{eqnarray}\label{eq33}
F^M=d(\upsilon-\varepsilon)+\left(\frac{\upsilon q_xk_y^2}{S}-\frac{\upsilon-\varepsilon}{q_x}\right)\sin\beta\cos\beta,
\end{eqnarray}
\begin{eqnarray}\label{eq34}
G^M=\left(\varepsilon-\frac{\upsilon k_y^2\left(\varepsilon\upsilon-k_x^2\right)}{S}\right)\frac{\sin^2\beta}{k_x},
\end{eqnarray}
and
\begin{eqnarray}\label{eq35}
\gamma_\upsilon^{HSL}=\frac{\upsilon^2\sigma_\upsilon^2Z}{2k_x^2q_x^4\sin^2\eta}\left(F^H\sin\alpha+G^H\cos\alpha\right)^2,
\end{eqnarray}
with
\begin{eqnarray}\label{eq36}
Z=\left(\Delta^2+k_y^2\right),
\end{eqnarray}
\begin{eqnarray}\label{eq37}
&& F^H=\upsilon(\upsilon-\varepsilon)d-\frac{\varepsilon\upsilon-k_x^2}{q_x}\sin\beta\cos\beta,\nonumber\\
&& G^H=k_x\sin^2\beta.
\end{eqnarray}
Expression~(\ref{eq35}) takes the simple form in the limiting case of $\delta$-function barriers (that is when $d\to0$, $\upsilon\to\infty$ but such that their product remains constant $\upsilon d=\varphi$)
\begin{eqnarray}\label{eq38}
\gamma_{\varphi}^{HSL}=\frac{\tilde{\sigma}_{\varphi}^2Z\sin^2\alpha}{2k_x^2\sin^2\eta},
\end{eqnarray}
with
\begin{eqnarray}\label{eq39}
\cos\eta=\cos\alpha\cos\varphi+\frac{\varepsilon}{k_x}\sin\alpha\sin\varphi.
\end{eqnarray}
Here $\varphi$ denotes the mean value of the ``potential'' at the $n$-th site: $\varphi_n=\varphi+\delta\varphi_n$, where $\delta\varphi_n$ are homogeneous random perturbations and $\tilde{\sigma}_{\varphi}^2=\langle\delta\varphi^2\rangle$.

In this case, the dependence of ILL on the barrier characteristics ($\upsilon$ and $d$) are determined only by means parameter $\varphi$ therefore, the formula~(\ref{eq38}) can also be obtained as a limit of the expression~(\ref{eq29}) provided that only barrier width fluctuates. It is obvious that in the limit of very narrow ($d\to0$) barriers superlattice with non-uniform gap becomes gapless and the equation~(\ref{eq38}) at $\Delta=0$ coincides with that of Ref.~\cite{Bliokh}
. The presence of a gap leads to localization of the particles incident on the structure at arbitrary angles, with the exception of delocalization resonances ($\alpha=\pi n$, $n=1,2,\dots$) that as in the case of gapless SL are exact for arbitrary disorder strength. When $\varphi=0$, that is for purely random $\delta$-potential Eq.~(\ref{eq38}) reduces to $\gamma_{\varphi}^{HSL}=\frac{\tilde{\sigma}_{\varphi}^2Z}{2k_x^2}$, which means complete localization of massive Dirac particles in such structure.

Now we return to the general case of a rectangular potential superlattices. As can be seen from the equation~(\ref{eq31}), for array with non-uniform gap all the states with $k_y\ne0$ (i.e., in the case of oblique incidence of the particles) are localized. When incidence angle $\theta_0=0$ perturbative delocalization resonances are determined vanishing expression in round brackets in Eq.~(\ref{eq31}), but unlike the geometric disorder, they exist only for weak disorder strength (Figs.~\ref{fig4}(a),~\ref{fig4}(b)). The weak-disorder expansion for the Lyapunov exponent for graphene SLs with uniform gap $\gamma_{\upsilon}^{HSL}$~(\ref{eq35}) also manifests an emergence of the delocalization resonances (at any angles of incidence) that occur under condition $F^H\sin\alpha+G^H\cos\alpha=0$ and disappear with increasing disorder strength (Figs.~\ref{fig4}(c),~\ref{fig4}(d)). Fig.~\ref{fig4} also clear demonstrates that although with the growth of disorder resonances disappear, but near them approximate formulas~(\ref{eq31}),~(\ref{eq35}) describes the ILL well even at a high degree of disorder $\delta=0.5$. Note that for a random potential without a regular superlattice component ($\upsilon_n=\delta\upsilon_n$) and for $k_y=0$ the Lyapunov exponent Eq.~(\ref{eq35}) reduced to
\begin{eqnarray}\label{eq40}
\gamma_{\upsilon}^{HSL}=\frac{\tilde{\sigma}_{\upsilon}^2\Delta^2\sin^2\beta}{2k_x^4}, \text{ } \tilde{\sigma}_{\upsilon}^2=\left\langle\delta\upsilon^2\right\rangle.
\end{eqnarray}
\begin{figure}[t] \centering
\includegraphics*[scale=0.6]{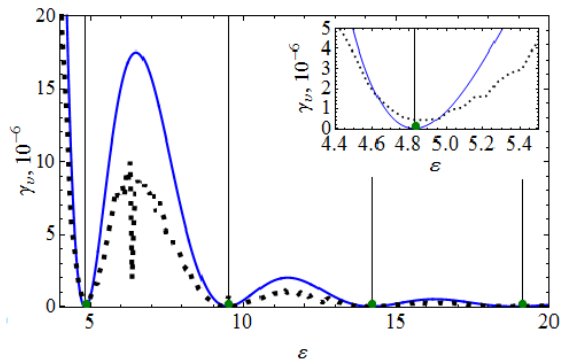}
\caption{Inverse localization length $\gamma_{\upsilon}^{HSL}$ at $\theta_0=0$, $\Delta=\pi/3$, $a=1/3$ and $d=2/3$ versus energy $\varepsilon$ for a purely random potential ($\upsilon=0$) with the amplitude of fluctuations $\delta\upsilon=\pi/10$. The analytical results [Eq.~(\ref{eq40})] and numerical data are shown by, respectively, the solid (blue) and dotted curves. Vertical lines mark the resonant energies.} \label{fig5}
\end{figure}
Unlike the case of $\delta$-function barriers, this expression vanishes for the resonance energies
\begin{eqnarray}\label{eq41}
\varepsilon_n=\pm\sqrt{\Delta^2+\left(\frac{\pi n}{d}\right)^2}, \text{ } n=1,2,\dots
\end{eqnarray}
leading to divergence of the localization length (Fig.~\ref{fig5}). This contradicts the statement of Ref.~\cite{Zhu} 
 that, in one-dimensional case massive Dirac particles should be localized for any weak disorders.

In the case where a fluctuating parameter is the gap value in the barrier region (and in the intervals between the barriers still $\Delta=0$), we find from Eqs.~(\ref{eq24}) -- (\ref{eq28})
\begin{eqnarray}\label{eq42}
&& \gamma_{\Delta}^{MSL}=\frac{\Delta^2\sigma_{\Delta}^2\sin^2\beta}{2q_x^2S}\nonumber\\
&& \cdot\left\{\frac{\Delta^2\sin^2\beta}{k_x^2q_x^2\sin^2\eta}\left(F_{\Delta}\sin\alpha+G_{\Delta}\cos\alpha\right)^2+\upsilon^2k_y^2\right\},
\end{eqnarray}
\begin{eqnarray}\label{eq43}
F_{\Delta}=\frac{\beta S-\left(\varepsilon\upsilon-k_x^2\right)^2\sin\beta\cos\beta}{k_xq_x\sin^2\beta},
\end{eqnarray}
\begin{eqnarray}\label{eq44}
G_{\Delta}=\varepsilon\upsilon-k_x^2.
\end{eqnarray}
Similarly to the previous case of fluctuating barrier height, the weak-disorder resonances are possible only for normally incident particles. Note that although these resonances obtained in the weak-disorder approximation they survive at significant deviations of the gap from its average value. Thus, numerical simulations carried out for the MSL with $\upsilon=\pi$, $\Delta=4\pi/9$ at $k_y=0$ showed that at $20\%$ of the gap fluctuation resonance remains well defined, although in the non-resonant zones there is a significant discrepancy between the approximate analytical (\ref{eq42}) and numerical results.

\section{Conclusion}

In summary, we have studied the localization behavior of Dirac particles in disordered graphene superlattices. Using the weak-disorder approximation, we obtained the analytical expression for the inverse localization length (Lyapunov exponent). The main attention was drawn to the two models of the SLs. One of them corresponds to the massive Dirac particles (the SL with homogeneous gap) in the presence of one-dimensional piecewise constant potentials. Another discussed model is a layered structure, made of gapped and gapless graphene strips. It is obvious, that the presence of a gapped graphene fraction in disordered SLs leads to suppression of the Klein tunneling and localization of Dirac particles with zero incidence angles. When the disorder emerges due to random thickness variations in the well (or barrier) layer the Fabry-Perot resonances leading to divergence of the localization length arise in the discussed SLs like in other models of the Kronig-Penney type. This result holds for the SLs, i.e. for infinite systems. When we consider a transmission probability through the lateral structure of finite length $L=Nl$, composed of alternating barrier and well strips (with the same or different gap magnitudes), the outer regions ($x<0$ and $x>L$) may correspond to both the well and the barrier parameters. Obviously, the transmittance will depend on these boundary conditions: $T_N$ -- for the well outer regions and ${T'}_N$ for the barrier ones. For example, when the width of wells fluctuate and resonance conditions $q_xd=\pi n$ are fulfilled, each of the barriers (and hence $N$ barriers) becomes completely transparent, i.e. $T_N=1$ for any $N$ and for any strength of disorder. At the same time, the equation $\gamma=0$ means that corresponding transmission probability through $N$ wells ${T'}_N$ cannot be an exponentially decaying function of $N$ (numerical data indicate that ${T'}_N$ does not decrease for a system made up of a sufficiently large number of layers). We also received the analytical expression for the localization length for the case of weakly fluctuating both barrier and well widths, taking into account possible correlations in disorder. We have studied and compared the cases where disorder is uncorrelated to cases where it is entirely correlated and anticorrelated. The main effect of the correlations, leading to an increase (or decrease) in the localization length, was found in the vicinities of double resonance arising under the conditions $\sin k_xa=0$, $\sin q_xd=0$.

Also, delocalization resonances for both types of the SL are obtained for the barriers with randomly varying height, but in contrast to the Fabry-Perot resonances, they are approximate. Resonance values of energy and angle of incidence are determined by the parameters of the system and, in general, can be found only numerically. Corresponding expressions for $\gamma_{\upsilon}$ ((\ref{eq31}),~(\ref{eq35})) demonstrate distinct features of two superlattice models: for massive Dirac particle resonance condition can be performed at arbitrary angles of incidence, while in the structures with non-uniform gap such weak-disorder delocalization is possible only for the Dirac particles with zero incidence angle. Interestingly, the delocalization states exist in one dimension (i.e. when $\theta_0=0$) for the massive Dirac particles with energies $\varepsilon_n$~(\ref{eq41}) placed in the purely random potential  i.e., with the barrier height  $\upsilon_n$ being a constant, randomly distributed in a certain range, which determines the degree of disorder. At the same time, for the disordered $\delta$-function potential without a regular superlattice component all the states are localized.

The obtained results for the localization length can be used for finding an analytical expression for the transmission coefficient through the finite-size disordered graphene system. In turn, it is possible to apply this expression to an analysis of conductance and resistance. However the analytical results were obtained by statistical averaging. Therefore, their comparison with corresponding experimental data requires sufficiently long structures. Despite this limitation, such a comparison for a single-mode microwave waveguide composed of relatively small number of cells ($N=26$) has shown quite good agreement in the case of weak disorder for the whole range of frequencies~\cite{Luna}
. In graphene heterostructures, a finiteness (or rather smallness) of mean free path and phase coherence length is another limiting factor for the emergence of sizable disordered superlattice effects (in particular the delocalization Fabry-Perot resonances). Accordingly, samples with high carriers’ mobility and small superlattice period are needed.

\section{Acknowledgments}

The authors thank A.M.~Satanin and A.~Konakov for interesting discussion. We also are grateful to V.A.~Burdov for his interest in this investigation and for helpful remarks. E.S.A. acknowledges support by the ``Dynasty'' Foundation.

\appendix
\section{}

To study the effect of disorder correlations we allow the width of barriers as well as the distance between them fluctuate relative to their average values, adding to the relations~(\ref{eq22}) correlator $\left\langle\rho_n^a\rho_{n'}^d\right\rangle$:
\begin{eqnarray}\label{eqa1}
&& \left\langle\rho_n^{a,d}\right\rangle=0, \text{ } \left\langle\rho_n^d\rho_{n'}^d\right\rangle=\sigma_d^2\delta_{nn'},\nonumber\\
&&\left\langle\rho_n^a\rho_{n'}^a\right\rangle=\sigma_a^2\delta_{nn'}, \text{ } \left\langle\rho_n^a\rho_{n'}^d\right\rangle=\sigma_{ad}\delta_{nn'}.
\end{eqnarray}
Accordingly the weak-disorder expansion of the transfer-matrix $\hat{M}_n$ is
\begin{eqnarray}\label{eqa2}
&& \hat{M}_n=\hat{M}+\alpha\hat{M}_{\alpha}'\rho_n^a+\beta\hat{M}_{\beta}'\rho_n^d+\frac{\alpha^2}{2}\hat{M}_{\alpha\alpha}''\left(\rho_n^a\right)^2 \nonumber\\
&& +\frac{\beta^2}{2}\hat{M}_{\beta\beta}''\left(\rho_n^d\right)^2+\alpha\beta\hat{M}_{\alpha\beta}''\rho_n^a\rho_n^d.
\end{eqnarray}
Here the matrix $\hat{M}$~(\ref{eq4}) and derivatives of $\hat{M}$ with respect to the fluctuating superlattice dimensions ($\hat{M}'$ and $\hat{M}''$) determined by the parameters of unperturbed SL. Following Ref.~\cite{Zhao}
, we find the matrix element $(P_N)_{11}$, using the representation ($M$-representation) in which the matrix $\hat{M}$ is diagonal $\hat{M}\to\tilde{\hat{M}}=$diag$(\lambda_+,\lambda_-)$
\begin{eqnarray}\label{eqa3}
&& \left(\tilde{P}_N\right)_{11}=\lambda_+^N\Biggl\{1\frac{1}{\lambda_+}\sum_{k=1}^N\Bigl[\alpha\tilde{M}_{\alpha}'\rho_k^a+\beta\tilde{M}_{\beta}'\rho_k^d \nonumber\\
&& +\frac{\alpha^2}{2}\tilde{M}_{\alpha\alpha}''\left(\rho_k^a\right)^2+\frac{\beta^2}{2}\tilde{M}_{\beta\beta}''\left(\rho_k^d\right)^2\nonumber\\
&& +\alpha\beta\tilde{M}_{\alpha\beta}''\rho_k^a\rho_k^d\Bigl]\Biggl\}.
\end{eqnarray}
Note that in this expression, we have omitted the terms proportional $\rho_k\rho_{k'\ne k}$, that vanish in the subsequent averaging.  In order to evaluate the disorder-induced Lyapunov exponent one need to combine Eqs.~(\ref{eq1}), (\ref{eq6}) and (\ref{eqa3}) and expand the logarithm within the quadratic approximation in the perturbation parameters. Performing averaging with help of Eq.~(\ref{eqa1}), we obtain

\begin{eqnarray}\label{eqa4}
&& \gamma=\frac{1}{2}\Biggl\{\alpha^2\sigma_a^2\Biggl[\left|\left(\tilde{M}_{\alpha}'\right)_{11}\right|^2+\text{Re}\frac{\left(\tilde{M}_{\alpha}''\right)_{11}}{\lambda_+}
\nonumber\\
&& -2\text{Re}^2\frac{\left(\tilde{M}_{\alpha}'\right)_{11}}{\lambda_+}\Biggl]+\beta^2\sigma_d^2\Biggl[\left|\left(\tilde{M}_{\beta}'\right)_{11}\right|^2
\nonumber\\
&& +\text{Re}\frac{\left(\tilde{M}_{\beta}''\right)_{11}}{\lambda_+}-2\text{Re}^2\frac{\left(\tilde{M}_{\beta}'\right)_{11}}{\lambda_+}\Biggl]\nonumber\\
&& +2\alpha\beta\sigma_{ad}\Biggl[\text{Re}\frac{\left(\tilde{M}_{\alpha\beta}''\right)_{11}}{\lambda_+}-\text{Im}\frac{\left(\tilde{M}_{\alpha}'\right)_{11}}{\lambda_+}\nonumber\\
&& \cdot\text{Im}\frac{\left(\tilde{M}_{\beta}'\right)_{11}}{\lambda_+}\Biggl]\Biggl\}.
\end{eqnarray}
Here the terms, proportional to $\alpha^2$ and $\beta^2$, lead to the expressions for $\gamma_a$ and $\gamma_d$ defined above by Eq.~(\ref{eq29}) without taking into account the correlation term ($\sim\alpha\beta$). To find it, we need to know $\frac{\tilde{M}_{11}'}{\lambda_+}$ and $\frac{\tilde{M}_{11}''}{\lambda_+}$. Computing respective derivatives of the matrix $M$ and making the transformation to the $M$-representation, after some algebraic calculations, we have
\begin{eqnarray}\label{eqa5}
\frac{\left(\tilde{M}_{\alpha(\beta)}'\right)_{11}}{\lambda_+}=-i\frac{\text{Re}a_{\alpha(\beta)}'}{\sin\eta},
\end{eqnarray}
\begin{eqnarray}\label{eqa6}
\frac{\left(\tilde{M}_{\alpha\beta}''\right)_{11}}{\lambda_+}=\frac{e^{-i\eta}\text{Re}\left(a_{\alpha}'^*a_{\beta}'-b_{\alpha}'^*b_{\beta}'\right)+\text{Re}a_{\alpha\beta}''}{i\sin\eta},
\end{eqnarray}
where $a$ and $b$ denote matrix elements of matrix $\hat{M}$ (Eq.~(\ref{eq4})): $a=M_{11}$, $b=M_{12}$. Using Eqs.~(\ref{eq14}),~(\ref{eq15}), we get from~(\ref{eqa5}) and~(\ref{eqa6}) the expressions that define the correlation term
\begin{eqnarray}\label{eqa7}
&& \text{Im}\left(\frac{\left(\tilde{M}_{\alpha}'\right)_{11}}{\lambda_+}\right)\text{Im}\left(\frac{\left(\tilde{M}_{\beta}'\right)_{11}}{\lambda_+}\right)=\frac{\bigl(\sin\alpha\cos\beta}{\sin^2\eta}\nonumber\\
&& =\frac{-f\cos\alpha\sin\beta\bigl)\left(-\sin\beta\cos\alpha+f\cos\beta\sin\alpha\right)}{\sin^2\eta},
\end{eqnarray}
\begin{eqnarray}\label{eqa8}
\text{Re}\left(\frac{\left(\tilde{M}_{\alpha\beta}''\right)_{11}}{\lambda_+}\right)=f.
\end{eqnarray}
Next, substituting Eqs.~(\ref{eqa7}), (\ref{eqa8}) into (\ref{eqa4}) we obtain the contribution of the correlation term $\Delta\gamma_{cor}$ in the inverse localization length
\begin{eqnarray}\label{eqa9}
\Delta\gamma_{cor}=\alpha\beta\sigma_{ad}\frac{\left(1-f^2\right)\sin\alpha\sin\beta\cos\eta}{\sin^2\eta},
\end{eqnarray}
where the function $f=f\left(\varepsilon,\theta_0\right)$ for both types of the considered SLs is defined be Eq.~(\ref{eq20}).


\begin{thebibliography}{99}

\bibitem{Park}
C.-H.~Park, L.~Yang, Y.-W.~Son, M.L.~Cohen, and S.G.~Louie, Nat. Phys. {\bf 4}, 213 (2008).

\bibitem{Brey}
L.~Brey and H.A.~Fertig, Phys. Rev. Lett. {\bf 103}, 046809 (2009).

\bibitem{Barb}
M.~Barbier, P.~Vasilopoulos, and F.M.~Peeters, Phys. Rev. B {\bf 81}, 075438 (2010).

\bibitem{Dell}
L.~Dell'Anna and A.~De~Martino, Phys. Rev. B {\bf 79}, 045420 (2009).

\bibitem{Le}
V.Q.~Le, C.H.~Pham, and V.L.~Nguyen, J. Phys. Cond. Matt. {\bf 24}, 345502 (2012).

\bibitem{Arov}
D.P~Arovas, L.~Brey, H.A.~Fertig, E.-A.~Kim, and K.~Zeigler, New J. Phys. {\bf 12}, 123020 (2010).

\bibitem{Esm1}
Mahammad~Esmailpour, Ayoub~Esmailpour, Reza~Asgari, M.~Elahi, and M.R.~Rahimi Tabar, Solid State Commun. {\bf 150}, 655 (2010).

\bibitem{Dub}
S.~Dubey, V.~Singh, A.K.~Bhat, P.~Parikh, S.~Grover, R.~Sensarma, V.~Tripathi, K.~Sengupt, M.M.~Deshmukh, Nano Lett. {\bf 13}, 3990 (2013).

\bibitem{Wang}
Li-Gong~Wang and Xi~Chen, J. Appl. Phys. {\bf 109}, 033710 (2011).

\bibitem{Son}
Y.-W.~Son, M.L.~Cohen, and S.G.~Louie, Phys. Rev. Lett. {\bf 97}, 216803 (2006).

\bibitem{Han}
M.Y.~Han, B.~\"{O}zyilmaz, Y.~Zhang, and P.~Kim, Phys. Rev. Lett. {\bf 98}, 206805 (2007).

\bibitem{Yan}
Q.~Yan, B.~Huang, J.~Yu, F.~Zheng, J.~Zang, J.~Wu, B.-L.~Gu, F.~Liu, and W.~Duan, Nano Lett. {\bf 7}, 1469 (2007).

\bibitem{Apel}
W.~Apel, G.~Pal, and L.~Schweitzer, Phys. Rev. B {\bf 83}, 125431 (2011).

\bibitem{Gui}
G.~Gui, J.~Li, and J.~Zhong, Phys. Rev. B {\bf 78}, 075435 (2008).

\bibitem{Pereira}
V.M.~Pereira, A.H.~Castro~Neto, and N.M.R.~Peres, Phys. Rev. B {\bf 80}, 045401 (2009).

\bibitem{Rib}
R.M.~Ribeiro, N.M.R.~Peres, J.~Coutinho, and P.R.~Briddon, Phys. Rev. B {\bf 78}, 075442 (2008).

\bibitem{Giov}
G.~Giovanetti, P.A.~Khomyakov, G.~Brocks, P.J.~Kelly, and J.~van~den~Brink, Phys. Rev. B {\bf 76}, 073103 (2007).

\bibitem{Zhou}
S.Y.~Zhou, G.-H.~Gweon, A.V.~Fedorov, P.N.~First, W.A.~de~Heer, D.-H.~Lee, F.~Guinea, A.H.~ Castro Neto, and A.~Lanzara, Nature Mat. {\bf 6}, 770 (2007).

\bibitem{Xiao}
Di~Xiao, Gui-Bin~Liu, W.~Feng, X.~Xu, and Wang~Yao, Phys. Rev. Lett. {\bf 108}, 196802 (2012).

\bibitem{Li}
Xiao~Li, Fan~Zhang, and Qian~Niu, Phys. Rev. Lett. {\bf 110}, 066803 (2013).

\bibitem{Peres}
N.M.R.~Peres, J. Phys. Cond. Matt. {\bf 21}, 095501 (2009).

\bibitem{Gomes}
Viana~Gomes and N.M.R.~Peres, J. Phys. Cond. Matt. {\bf 20}, 325221 (2008).

\bibitem{Giav}
G.~Giavaras, and F.~Nori, Appl. Phys. Lett. {\bf 97}, 243106 (2010); Phys. Rev. B {\bf 83}, 165427 (2011).

\bibitem{Ratn}
P.V.~Ratnikov and A.P.~Silin, Phys. Solid State {\bf 52}, 1763 (2010).

\bibitem{Maks}
G.M.~Maksimova, E.S.~Azarova, A.V.~Telezhnikov, and V.A.~Burdov, Phys. Rev. B {\bf 86}, 205422 (2012).

\bibitem{Azar}
E.S.~Azarova and G.M.~Maksimova, Physica E {\bf 61}, 118 (2014).

\bibitem{Fert}
H.A.~Fertig and L.~Brey, Phys. Trans. R. Soc. {\bf 368}, 5483 (2010).

\bibitem{Nom}
K.~Nomura, M.~Koshino, and S.~Ruy, Phys. Rev. Lett. {\bf 99}, 146806 (2007).

\bibitem{Zhu}
Shi-Liang~Zhu, DAn-Wel~Zhang, and Z.D.~Wang, Phys. Rev. Lett. {\bf 102}, 210403 (2009).

\bibitem{Bliokh}
Yury~P.~Bliokh, Valentin Freilikher, Sergey Savel'ev, and Franco Nori, Phys. Rev. B {\bf 79}, 075123 (2009).

\bibitem{Abed}
N.~Abedpour, Ayoub Esmailpour, Reza Asgari, and M.Reza~Rahimi~Tabar, Phys. Rev. B {\bf 79}, 165412 (2009).

\bibitem{Zhao}
Qifang~Zhao, Jiangbin~Gong, and Cord~A.~M\"{u}ller, Phys. Rev. B {\bf 85}, 104201 (2012).

\bibitem{Esm2}
Ayoub~Esmailpour, Fatemeh~Pakdel, and Razieh~Jahanaray, Physica E {\bf 54}, 214 (2013).

\bibitem{Mark}
P.~Marco\v{s} and Costas~M.~Soukoulis, Wave Propagation. From Electrons to Photonic Cristalls and Laft-Handed Materials, Princeton University Press, Princeton (2008).

\bibitem{Kats}
M.I.~Katsnelson, K.S.~Novoselov, and A.K.~Geim, Nature Phys. {\bf 2}, 620 (2006).

\bibitem{Milt}
J.~Milton~Pereira~Jr., P.~Vasilopoulos, and F.M.~Peeters, Appl. Phys. Lett. {\bf90}, 132122, (2007).

\bibitem{Shytov}
A.V.~Shytov, M.S.~Rudner, and L.S.~Levitov, Phys. Rev. Lett. {\bf101}, 156804, (2008).

\bibitem{Masir}
M.~Ramezani~Masir, P.~Vasilopoulos, and F.M.~Peeters, Phys. Rev. B {\bf82}, 115417, (2010).

\bibitem{Luna}
G.A.~Luna-Acosta, F.M.~Izrailev, N.M.~Makarov, U.~Kuhi, and H.-J.~St\"{o}ckmann, Phys. Rev. B {\bf 80}, 115112 (2009).

\bibitem{Mog}
D.~Mogilevtsev, F.A.~Pinheiro, R.R~dos~Santos, S.B.~Cavolcanti, and L.E.~Oliveira, Phys. Rev. B {\bf 84}, 094204 (2011).

\end{thebibliography}
\end{document}